\documentclass[12pt]{article}

\usepackage{amsmath,amsfonts,amssymb,amsthm,color,hyperref,url,geometry,setspace,natbib,ulem,appendix}
\hypersetup{
    colorlinks=true,
    linkcolor=blue,
    filecolor=magenta,      
    urlcolor=cyan,
    citecolor=blue,
}
\usepackage{graphicx,booktabs,multirow}
\usepackage{algorithm,algcompatible}
\usepackage[noend]{algpseudocode}
\numberwithin{equation}{section}
\usepackage{graphicx,booktabs,multirow,tabularx}
\usepackage{algorithm,algpseudocode}
\usepackage{pdfsync}
\usepackage{float,subfig}
\usepackage{tikz}
\usepackage{natbib}
\usepackage{accents}
\usepackage{xr}
\graphicspath{{./figures/}}

\algnewcommand{\Input}{\textbf{Input: }}
\algnewcommand{\Init}{\textbf{Initilization: }}
\algnewcommand{\Output}{\textbf{Output: }}

\newcommand{\norm}[1]{\|#1\|}

\newcommand{\wh}[1]{\widehat{#1}}
\newcommand{\wt}[1]{\widetilde{#1}}
\newcommand{\rank}[1]{\text{rank}(#1)}
\newcommand{\diag}[1]{\text{diag}\{#1\}}
\newcommand{\trace}[1]{\text{trace}\{#1\}}
\newcommand{\abs}[1]{\vert #1 \vert}
\newcommand{\inner}[1]{\langle#1\rangle}
\newcommand{\op}{\text{op}}

\def\trans{^{\top}}

\def\strans{^{*\top}}

\def\code#1{\texttt{#1}}

\newtheorem{condition}{Condition}

\newcommand{\bTheta}{\boldsymbol{\Theta}}
\newcommand{\bSigma}{\boldsymbol{\Sigma}}

\newcommand{\bbeta}{\boldsymbol{\beta}}
\newcommand{\balpha}{\boldsymbol{\alpha}}
\newcommand{\btheta}{\boldsymbol{\theta}}

\newcommand{\bmu}{\boldsymbol{\mu}}

\newcommand{\mcL}{\mathcal{L}}

\def\bX{{\bf X}}
\def\bY{{\bf Y}}
\def\bU{{\bf U}}
\def\bV{{\bf V}}
\def\bD{{\bf D}}
\def\bI{{\bf I}}
\def\bE{{\bf E}}

\def\bA{{\bf A}}
\def\bB{{\bf B}}
\def\bW{{\bf W}}
\def\bL{{\bf L}}
\def\bS{{\bf S}}
\def\bQ{{\bf Q}}

\def\bM{{\bf M}}

\def\bZ{{\bf Z}}

\def\bO{{\bf O}}

\def\mbR{\mathbb{R}}
\def\mbE{\mathbb{E}}
\def\mbG{\mathbb{G}}
\def\mbP{\mathbb{P}}
\def\mbB{\mathbb{B}}

\def\x{\boldsymbol{x}}
\def\y{\boldsymbol{y}}
\def\e{\boldsymbol{e}}
\def\c{\boldsymbol{c}}

\def\v{\boldsymbol{v}}
\def\z{\boldsymbol{z}}
\def\a{\boldsymbol{a}}
\def\b{\boldsymbol{b}}

\def\m{\boldsymbol{m}}

\def\0{\boldsymbol{0}}
\def\1{\boldsymbol{1}}

\newtheorem{proposition}{Proposition}
\newtheorem{remark}{Remark}
\newtheorem{theorem}{Theorem}
\title{Generalized Reduced-Rank Regression with Homogeneity Pursuit}

\usepackage[affil-it]{authblk} 
\usepackage{etoolbox}
\usepackage{lmodern}

\makeatletter
\patchcmd{\@maketitle}{\LARGE \@title}{\fontsize{16}{19.2}\selectfont\@title}{}{}
\makeatother

\author[1]{Ruipeng Dong}
\author[2]{Ganggang Xu}
\author[3]{Yongtao Guan}
\affil[1]{International Institute of Finance, School of Management, University of Science and Technology of China}
\affil[2]{Department of Management Science, University of Miami}
\affil[3]{School of Data Science, The Chinese University of Hong Kong, Shenzhen}

\date{}

\begin{document}
  \maketitle

  \begin{abstract}
    Homogeneity, low rank, and sparsity are three widely adopted assumptions in multi-response regression models to address the curse of dimensionality and improve estimation accuracy. However, there is limited literature that examines these assumptions within a unified framework. In this paper, we investigate the homogeneity, low rank, and sparsity assumptions under the generalized linear model with high-dimensional responses and covariates,  encompassing a wide range of practical applications. Our work establishes a comprehensive benchmark for comparing the effects of these three assumptions and introduces a regularized maximum likelihood estimation method to fit the corresponding models. Under mild conditions,we prove the statistical consistency of our estimator. Theoretical results provide insights into the role of homogeneity and offer a quantitative analysis of scenarios where homogeneity improves estimation accuracy.  The proposed method’s effectiveness is demonstrated through numerical simulations and an empirical analysis of tree species data from Barro Colorado Island.
    \smallskip 

  \end{abstract}

  \section{Introduction}


The integrative learning with multiple responses is a prevalent challenge in finance, operations, marketing, social media, and other fields \citep{mcfadden1977modelling,geweke1996bayesian,chen2015recent,zhou2013learning,hessellund2022semiparametric,hessellund2022second}. Multi-response regression is a widely adopted method for addressing this issue, where the model is parameterized by a coefficient matrix. However, estimating the model becomes increasingly difficult when there are too many responses due to limited sample sizes. For instance, in the conjoint analysis of markets \citep{kamakura2012menu,eggers2021choice}, firms need to estimate a matrix to characterize consumer preferences based on a finite number of observations, with each consumer represented by a corresponding column. This leads to high-dimensional response data. Due to the curse of dimensionality, estimation accuracy deteriorates as the number of responses increases. To obtain accurate estimates, \citet{evgeniou2007convex} and \citet{chen2017modeling} assume that consumer types can be grouped into a limited number of segments, effectively clustering the columns of the matrix during estimation. This approach is based on the assumption of column homogeneity. Similarly, \citet{price2018cluster} utilize column homogeneity to enhance the estimation and prediction accuracy of multi-response regression models, thereby addressing the curse of dimensionality. Building on this idea, \citet{sherwood2024use} propose incorporating minimum penalties to simultaneously estimate regression coefficients while capturing homogeneous relationships among multiple responses. For a comprehensive review of related methodologies, see \citet{price2022detecting}.


In addition to the homogeneity assumption, the low-rank assumption is widely employed to achieve accurate estimation and prediction when dealing with high-dimensional responses. Under this framework, researchers assume that the coefficient matrix has a low rank, leading to the development of reduced-rank regression (RRR) \citep{reinsel2022multivariate}. Beyond the challenge of high-dimensional responses, the rise of data science has introduced regression models with massive numbers of covariates, many of which are irrelevant and hinder accurate model estimation. To address this issue, variable selection methods have been developed, incorporating sparsity assumptions on the coefficients to enhance model interpretability and identifiability \citep{fan2010selective}. Combining sparsity with RRR, sparse reduced-rank regression (SRRR) has been introduced, where most rows of the low-rank coefficient matrix are zero, effectively excluding irrelevant covariates from the model. A growing body of literature has explored the estimation of sparse and low-rank matrices from both statistical and computational perspectives, employing regularization techniques and matrix factorization to induce these structures \citep{chen2012sparse,bunea2012joint,uematsu2019sofar,chen2022fast,bunea2011optimal,chen2013reduced,sun2016guaranteed,ge2017no,zhu2023simultaneous,fang2024group,liu2024two}.


While numerous studies have examined homogeneity, low rank, and sparsity in multi-response models, few have addressed these assumptions within a unified framework. Note that matrix homogeneity implies that the matrix can be approximated, or even equal, to a low-rank matrix. For instance, if all columns are identical, indicating they belong to a single group, the matrix rank is one. Therefore, it is essential to integrate homogeneity and low rank into a unified model. Existing literature typically focuses on either homogeneity or the combination of sparsity and low rank. Recently, \citet{she2022supervised} investigated the homogeneity of rows in reduced-rank regression, assuming that the rows of a low-rank matrix can be clustered into a few groups, which we refer to as row homogeneity. Although their work incorporates both low rank and homogeneity, it assumes exact equality among rows within the same group, potentially leading to model misspecification when rows are only approximately equal. Furthermore, row homogeneity is not suitable for clustering responses, a common requirement in applications such as conjoint analysis. Therefore, it is necessary to study the homogeneity of columns, low rank, and sparsity assumptions within a unified framework to understand how within-class differences, the number of groups, and low rank influence estimation accuracy.

In this paper, we investigate the generalized linear model with high-dimensional responses and covariates, focusing on the column homogeneity, low rank, and sparsity within a unified framework. Specifically, we employ a regularized maximum likelihood estimation (MLE) approach, incorporating an $\ell_2$-norm fusion penalty to facilitate column clustering. The coefficient matrix is factorized into the product of two smaller matrices, ensuring that the low-rank constraint is inherently satisfied. Additionally, sparsity is induced by constraining the number of nonzero rows in the coefficient matrix. Our contributions are threefold. First, the proposed method encompasses a broad class of models and integrates the assumptions of column homogeneity, low rank, and sparsity into a unified framework. This approach does not require these assumptions to hold simultaneously but provides compatibility with them, establishing a consistent benchmark for comparing their respective effects. Second, we introduce an iterative block gradient descent algorithm to solve the regularized MLE problem, with theoretical guarantees on its convergence. Finally, our theoretical analysis elucidates how and when column homogeneity leads to improved estimation accuracy and offers a quantitative assessment of its impact on statistical consistency. Moreover, the proposed method remains flexible even when the columns do not satisfy the homogeneity assumption; in such cases, the estimation and corresponding statistical properties revert to those of SRRR model. 


The remainder of this paper is structured as follows. Section \ref{sec:model} outlines the model setup and proposed methodology, along with the detailed algorithm. Theoretical analysis is provided in Section \ref{sec:theory}. Lastly, Section \ref{sec:simulation} presents numerical simulations and an empirical study on tree species data from Barro Colorado Island. The technical proofs are provided in the appendix and online supplementary materials for reference.

\smallskip 

\noindent\textbf{Notations.} Throughout the paper, we use the bold lowercase and uppercase to denote the vector and matrix, respectively. For an arbitrary vector $\a = (a_j)\in\mbR^p$, we define the norm $\norm{\a}_q = (\sum_{j=1}^p \abs{a_j}^q)^{1/q}$ with $0 < q < +\infty$. For an arbitrary matrix $\bA$, we use $\norm{\bA}_F$ and $\norm{\bA}_{\op}$ to denote the Frobenius and operator norms, respectively, and $\norm{\bA}_{2,0}$ to denote the number of nonzero rows of $\bA$. Moreover, let $\rank{\bA}$ be the rank of $\bA$. 

\section{Problem Formulation}\label{sec:model}

\subsection{Basic model setups}\label{subsec:model}

In this section, we present the model setups, speficially focusing on the multinomial logistic regression model and the generalized linear model. While these models share similar properties, they differ slightly in their link functions. To ensure clarity, we will introduce each model separately in the following subsections.

\subsubsection{Multinomial Logistic Model}

Suppose we have $(m+1)$ categories and make the last as the reference category. 
Moreover, we define the $i$-th response vector as $\y_i=[y_{i1},\dots,y_{im}]\in\mbR^m$, where $y_{ij}$ is equal to 1 if the observation $i$ belongs to the category $j$, otherwise $y_{ij}$ is zero. Then the distribution of $\y_i$ is defined as follows
\begin{equation}
    \mbP(y_{ij} = 1) = \frac{\exp(\theta_{ij}^*)}{1 + \sum_{k=1}^m \exp(\theta_{ik}^*)} ~ 
    \text{with}~ j = 1,2,\dots,m, \label{eq:mnl:model}
\end{equation}
where the natural parameter $\theta_{ij}^*$ is modeled as $\theta_{ij}^* = \z_i\trans\balpha_j^* + \x_i\trans\bbeta_j^*$, with covariates $\z_i \in \mbR^q$ and $\x_i \in \mbR^p$. For subsequent homogeneity analysis, we categorize the covariates into two types: (1) for all responses, the coefficients $\balpha_j^* \in \mbR^q$ associated with variables $\z_i \in \mbR^q$ are heterogeneous across different $j$'s; (2) the coefficients $\bbeta_j^* \in \mbR^p$ associated with variables $\x_i$ are homogeneous within the same group for homogeneity. Additionally, we assume that the covariate $\z_i$ includes an intercept term, where the first element of $\z_i$ is equal to $1$. The category of covariates includes cases where homogeneity does not exist as described in type (1) for the sake of generality. However, in this paper, we primarily focus on the homogeneous part as defined in type (2).

With the last category as the reference, we only need to estimate the first $m$ coefficient vectors $(\balpha_j^*, \bbeta_j^*)$. Given $n$ i.i.d. observations, the negative log-likelihood is 
\begin{align*}
    -\log\left(\prod_{i=1}^n \prod_{j=1}^m p_{ij}^{y_{ij}}\right)
    \propto \sum_{i=1}^n \log \left[1 + \sum_{j=1}^m \exp(\theta_{ij})\right] - \sum_{i=1}^n \sum_{j=1}^m y_{ij}\theta_{ij},
\end{align*}
where $p_{ij} = \frac{\exp(\theta_{ij})}{1+\sum_{j=1}^m \exp(\theta_{ij})}$ and $\theta_{ij} = \z_i\trans \balpha_j + \x_i\trans \bbeta_j$. For clarity, we define the following two coefficient matrices
\begin{align*}
    \bA = [\balpha_1, \dots, \balpha_m] \in \mbR^{q \times m} \quad \text{and} \quad \bB = [\bbeta_1, \dots, \bbeta_m] \in \mbR^{p \times m}.
\end{align*}
The true coefficients corresponding to the true distribution are denoted by $\bA^*$ and $\bB^*$. Let $\bX = [\x_1, \dots, \x_n]\trans \in \mbR^{n \times p}$ and $\bZ = [\z_1, \dots, \z_n]\trans \in \mbR^{n \times q}$ be the covariate matrices. With the natural parameter vector $\btheta_i = \bA\trans \z_i + \bB\trans \x_i = (\theta_{ij}) \in \mbR^m$, the negative log-likelihood can be rewritten as
\begin{align}
    \sum_{i=1}^{n} b(\btheta_i) - \trace{\bY\trans \bTheta}, \label{loss:1}
\end{align}
where $\bY = [\y_1,\dots,\y_n]\trans\in\mbR^{n\times m}$ and $b(\btheta_i) = \log[1 + \sum_{j=1}^m \exp(\theta_{ij})]$. Here $\bTheta = [\btheta_1, \dots, \btheta_n]\trans$ is a $n\times m$ matrix stacked by natural parameters, which also can be expressed as $\bTheta = \bZ\bA + \bX\bB$.

\subsubsection{Generalized Linear Model}

Following the same notations of the multinomial logistic model, we let $\y_i\in\mbR^m$ be the $i$-th response vector and each element of $\y_i$ follows a exponential dispersion distribution. Then the density function of $y_{ij}$ is defined as 
\begin{equation*}
    f_j(y_{ij};\theta_{ij}^*,\phi_j) = \exp\left\{\frac{y_{ij}\theta_{ij}^* - b_j(\theta_{ij}^*)}{a_j(\phi_j^*)} + c_j(y_{ij};\phi_j^*)\right\},
\end{equation*}
where $\theta_{ij}^*$ is the natural parameter of $y_{ij}$, $\phi_j$ is the dispersion parameter of the $j$-th response. Moreover, $a_j,b_j$ and $c_j$ are the known functions determined by the specific distribution of $y_{ij}$. Common distributions within the exponential dispersion family are summarized in Table \ref{tab:1}. The first derivative of $b_j(\theta_{ij})$ satisfies $\nabla b_j(\theta_{ij}^*) = \mbE(y_{ij})$. Incorporating the covariates $\z_i\in \mbR^q$ and $\x_i\in \mbR^p$, we model the natural parameter as $\theta_{ij}^* = \balpha_j\strans\z_i + \bbeta_j\strans\x_i$, consistent with the multinomial logistic model. Then the negative log-likelihood is given by
\begin{equation*}
    \begin{aligned}
        \sum_{i=1}^n\sum_{j=1}^m \left\{\frac{b_j(\theta_{ij}) - y_{ij}\theta_{ij}}{a_j(\phi_j)} - c_j(y_{ij};\phi_j)\right\}
        \propto  \sum_{i=1}^n\sum_{j=1}^m b_j(\theta_{ij}) - y_{ij}\theta_{ij}.
    \end{aligned}
\end{equation*}
Similar to the multinomial logistic model, we define the function $b(\btheta_i) = \sum_{j=1}^m b_j(\theta_{ij})$, where the specific forms of $b_j$ are listed in Table \ref{tab:1}. Consequently, the negative log-likelihood is same to the equation \eqref{loss:1} and we have that $\nabla b(\btheta_i^*) = \mbE(\y_i)$, a property also holding for the multinomial logistic model. 

\begin{table}[!htp]
    \centering
    \caption{Some common distributions in the exponential dispersion family. \label{tab:1}}
    \resizebox{\columnwidth}{!}{
        \begin{tabular}{l|c|c|c|c|c|c|c}
            \toprule
            Distribution & Mean      & Variance   & $\theta$        & $\phi_j$     & $a_j(\phi)$ & $b_j(\theta)$          & $c_j(y;\phi)$
            \\
            \hline
            Normal       & $\mu$     & $\sigma^2$ & $\mu$           & $\sigma^2$ & $\phi$    & $\theta^2/2$         & $-(y^2\phi^{-1}+\log(2\pi))/2$
            \\
            Bernoulli    & $p$       & $p(1-p)$   & $\log[p/(1-p)]$ & $1$        & $1$       & $\log(1+e^{\theta})$ & $0$
            \\
            Poisson      & $\lambda$ & $\lambda$  & $\log\lambda$   & $1$        & $1$       & $e^{\theta}$         & $-\log(y!)$
            \\
            \bottomrule 
        \end{tabular}
    }
\end{table}

\begin{remark}
    The primary distinction between the multinomial logistic model and the generalized linear model lies in the formula of the link function $\b(\cdot)$. In the generalized linear model, $\b(\cdot)$ can be represented as the sum of individual functions $b_j(\cdot)$, whereas the multinomial logistic model lacks this additive property. 
\end{remark}

\begin{figure*}
    \centering 
    \subfloat[Heatmap of $\bB^*$]{\includegraphics[scale=0.35]{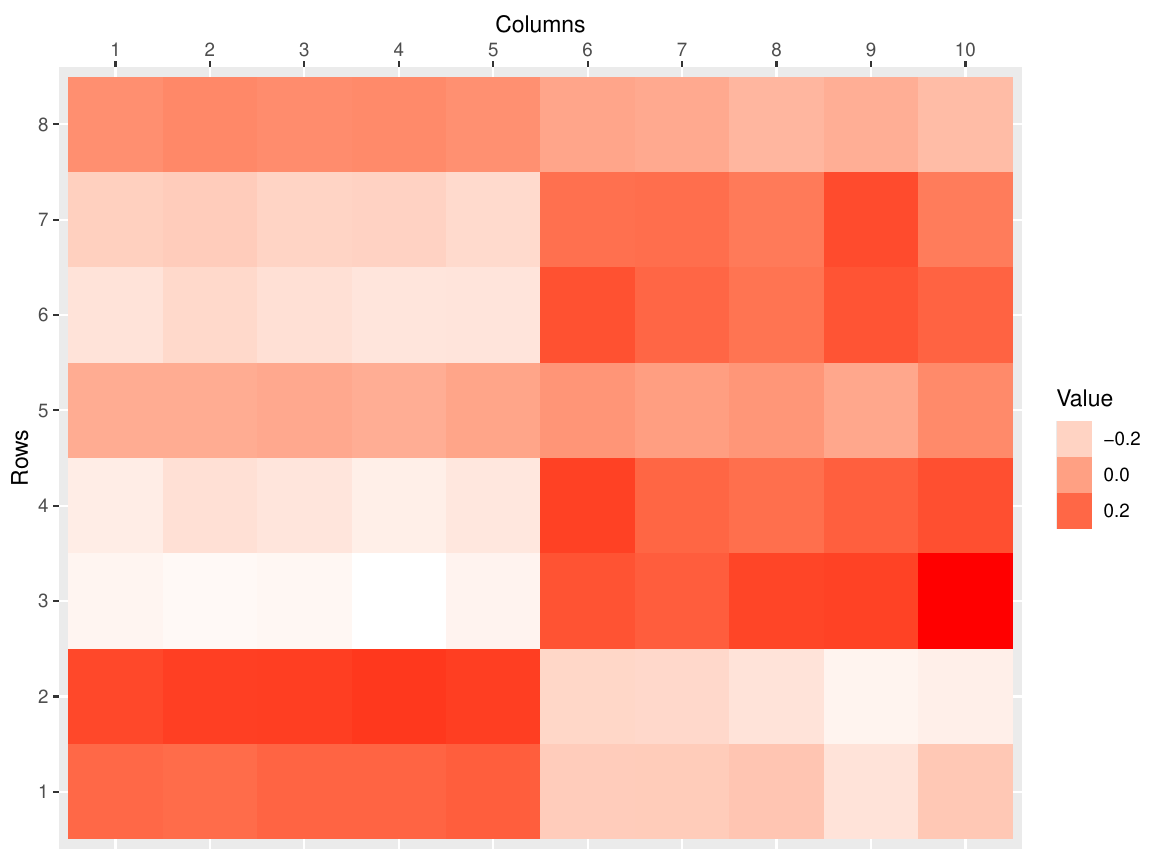}}
    \subfloat[Heatmap of $\bV^*\bD^*$]{\includegraphics[scale=0.35]{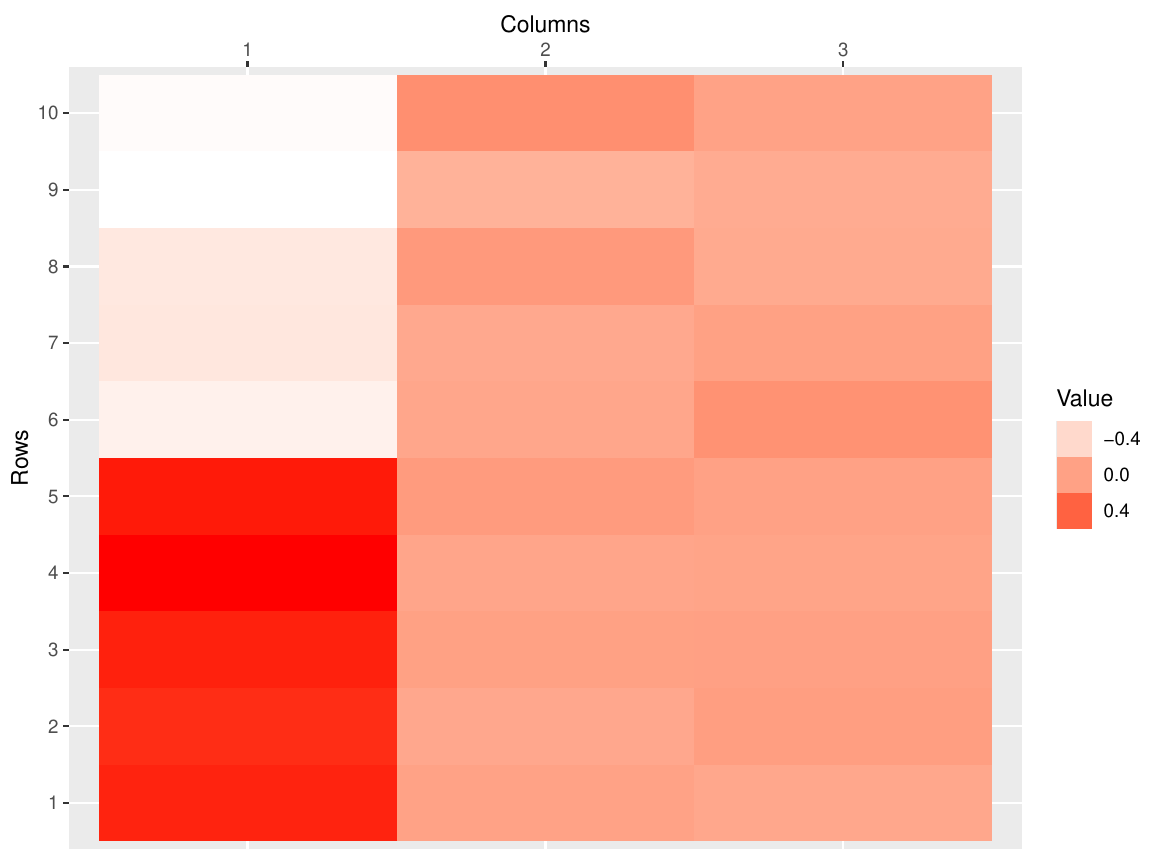}}
    \caption{An illustration of the homogeneity: the columns of $\bB^*$ can be clustered into two groups, columns 1-5 and columns 6-10. Let $\bU^*\bD^*\bV\strans$ be its SVD; thus, the columns of $\bB^*$ correspond to the rows of $\bV^*\bD^*$. Heatmaps reveal a significant group structure in both. Notably, the difference between groups is mainly reflected in the first column of $\bV^*\bD^*$, implying that homogeneity reduces the complexity of low-rank models. \label{fig:example}}
\end{figure*}

Recall that $\bbeta_j^*$ is the $j$-th column of $\bB^*$. In this paper, we assume that the true coefficient $\bbeta_j^*$ can be classified into $K^*$ groups, each coded by $1, \dots, K^*$, with $\mbG^* = (g_1^*, \dots, g_m^*)^\top$ indicating the group membership of $\bbeta_j^*$. The underlying reasoning behind such group assumptions is that the associations between certain response variables and covariates may exhibit similarities, which is reasonable in many applications. We further impose sparsity on the rows of $\bB^*$ and assume that $\bB^*$ is low-rank, where $\norm{\bB^*}_{2,0} = s^*$, $\rank{\bB^*} = r^*$, and $\max\{q, s^*, r^*\} \ll \min\{n, m, p\}$. Here, $q$ denotes the dimension of $\z_i$, a low-dimensional covariate vector. We then consider simultaneous clustering and variable selection through the estimation of matrix $\bB^*$.
To illustrate the homogeneity and low-rank assumptions, we plot an example on heatmaps of $\bB^*$ and its singular value decomposition (SVD) in Figure \ref{fig:example}, where $\bB^*$ is an $8\times 10$ matrix with $\rank{\bB^*}=3$. This example suggests that the homogeneity can reduce the complexity of low-rank model significantly.

\subsection{Homogeneity Pursuit with Reduced-Rank Structure} 

\subsubsection{Regularized negative likelihood}

In the above section, we introduced the setups of multinomial logistic and generalized linear models, along with the corresponding negative log-likelihood. Using the expression $\bTheta = \bZ\bA + \bX\bB$, we can reformulate the negative log-likelihood as 
\begin{align}
    \ell(\bA,\bB) \triangleq \sum_{i=1}^n b(\bA\trans\z_i + \bB\trans\x_i) - \sum_{i=1}^n \sum_{j=1}^m y_{ij}(\balpha_j\trans\z_i + \bbeta_j\trans\x_i). \label{eq:likelihood}
\end{align}
Before presenting the detailed algorithm, we first define some notations that will be used subsequently. For any matrix $\bB = [\bbeta_1, \dots, \bbeta_m] \in \mathbb{R}^{p \times m}$, let $\mbG = (g_1, \dots, g_m)^\top$ denote the vector of group memberships for the columns of $\bB$. The adjacency matrix $\bO(\mbG)$ is an $m \times m$ matrix where $o_{ij} = 1$ if $g_i = g_j$, indicating that columns $i$ and $j$ belong to the same group, and $o_{ii} = 0$ for all $i$. The degree matrix $\bD(\mbG) = \diag{d_1, \dots, d_m}$ is defined with $d_i = \sum_{j=1}^m o_{ij}$, and the unnormalized Laplacian matrix is given by $\wt{\bL} = \bD - \bO$. For simplicity, we omit $\mbG$ in the expressions of the adjacency and degree matrices subsequently. 

Additionally, let $G_k = \{j : g_j = k\}$ denote the $k$-th group. We define a diagonal matrix $\bW \in \mathbb{R}^{m \times m}$, where the $j$-th diagonal entry is $1/\abs{G_k}$ if $j \in G_k$, with $\abs{G_k}$ representing the size of group $G_k$. Consequently, the normalized Laplacian matrix $\bL = \bW^{-1/2} \wt{\bL} \bW^{-1/2}$ is used to cluster the columns of $\bB$, thereby identifying the column homogeneity within $\bB$.

Following the aforementioned definition, we proceed to introduce regularized likelihood estimation as detailed below
\begin{equation}
    \begin{aligned}
         & \underset{\bA,\bB,\mbG}{\min}\,\, \ell(\bA,\bB) + \frac{\lambda}{2} \trace{\bB\bL\bB\trans},
        \\
         & \quad \text{s.t.} \,\, \norm{\bB}_{2,0} \leq s ~ \text{and} ~ \rank{\bB} \leq r,
    \end{aligned} \label{opt:1}
\end{equation}
where the Laplacian matrix $\bL$ is determined by the membership indicator $\mbG$. Motivated by the low-rank structure of $\bB$, we re-parameterize $\bB$ as the product of two smaller matrices, $\bU\bV\trans$. Consequently, the optimization problem \eqref{opt:1} can be reformulated as
\begin{equation}
    \begin{aligned}
         & \underset{\bA,\bU,\bV,\mbG}{\min}\,\,
        \mathcal{L}(\bA,\bU,\bV,\mbG) \triangleq \ell(\bA,\bU\bV\trans) + \frac{\lambda}{2} \trace{\bU\bV\trans\bL\bV\bU\trans}
        \\
         & \quad \text{s.t.}\,\, \norm{\bU}_{2,0} \leq s ~\text{and}~ \bV\trans\bV = \bI_r.
    \end{aligned} \label{opt:2}
\end{equation}
The orthogonal constraint on $\bV \in \mbR^{m \times r}$ ensures that the sparsity constraint on $\norm{\bU}_{2,0}$ is equivalent to the original constraint on $\norm{\bB}_{2,0}$ because it preserves the $\ell_2$ norm of the rows between $\bU$ and $\bB$.

\subsubsection{Blockwise updates by majorizing function}\label{subsec:blockwise}

In this section, we employ the majorize-minimization framework to solve the optimization problem \eqref{opt:2}. The algorithm updates $\bA$, $\bU$, $\bV$, and $\mbG$ iteratively. For notational consistency, let $\nabla^2 b_j(\theta_{ij})$ denote the $j$-th diagonal element of the Hessian matrix $\nabla^2 b(\btheta_i) \in \mathbb{R}^{m \times m}$, where $b(\btheta)$ is the link function associated with the natural parameter as defined in Section \ref{subsec:model}, and the subscript $i$ indicates evaluation at the $i$-th observation. In addition, let $\bA^t$ and $\bB^t = \bU^t(\bV^t)\trans$ denote the solution at the $t$-th iteration. Consequently, we derive a surrogate function with quadratic loss, which is presented below.
\begin{proposition}\label{prop:4}
	For the negative log-likelihood $\ell(\bA, \bB)$, we have a surrogate function satisfies 
	\begin{align*}
		\ell(\bA, \bB) \leq \frac{w_t}{2} \norm{\bY^t - \bZ\bA - \bX\bB}_F^2 + C_t,
	\end{align*}
	where $C_t$ is a constant determined by $(\bA^t, \bB^t)$. As for $w_t$, we have $w_t = \max_{i,j} 2\nabla^2 b_j(\theta_{ij}^t)$ under the multinomial logistic model. Otherwise, $w_t = \max_{i,j} \nabla^2 b_j(\theta_{ij}^t)$ under the generalized linear model. Moreover, $\bY^t$ is a $n\times m$ artificial response matrix, whose $i$-th row is defined as $\y_i^t = \btheta_i^t + (\y_i - \nabla b(\btheta_i^t)) / w_t$. 
\end{proposition}

By substituting the likelihood loss with a surrogate function and replacing $\bB$ with $\bU\bV\trans$, we derive a surrogate objective function for the optimization presented in \eqref{opt:2} as follows
\begin{equation}
    \begin{aligned}
         & \underset{\bA,\bU,\bV,\mbG}{\min} ~ L_t(\bA,\bU,\bV,\mbG) \triangleq \frac{w_t}{2} \norm{\bY^t - \bZ\bA - \bX\bU\bV\trans}_F^2 + \frac{\lambda}{2} \trace{\bU\bV\trans\bL\bV\bU\trans}
        \\
         & \quad \text{s.t.} ~ \norm{\bU}_{2,0} \leq s ~ \text{and} ~ \bV\trans\bV = \bI_r,
    \end{aligned} \label{opt:4}
\end{equation}
where the subscript $t$ in $L_t$ indicates that we majorize the original objective function at the $t$-th iteration, specifically at $(\bA^t, \bU^t, \bV^t, \mbG^t)$. The surrogate optimization \eqref{opt:4} allows us to update $(\bA, \bU, \bV, \mbG)$ analytically, and the detailed procedure is outlined below. 

\smallskip

\noindent\textbf{(a) Update of $\bU$}. 

We update the matrix $\bU$ using gradient descent combined with a hard-thresholding rule, a method that has been extensively utilized in optimization problems involving $\ell_0$-norm constraints \citep{jain2014iterative}. In details, the update is as follows
\begin{equation}
    \wt{\bU}^{t+1} \leftarrow \bU^t - \eta \nabla L_t (\bU^t), \quad
    \bU^{t+1} \leftarrow H_s(\wt{\bU}^{t+1}), \label{update:u}
\end{equation}
where $\eta$ is a pre-specified step size, and $H_s(\cdot)$ denotes the hard-thresholding operator that selects the first $s$ rows with the largest $\ell_2$ norms. In addition, the gradient $\nabla L_t (\bU^t)$ is 
\begin{align*}
    \nabla L_t = w_t\left[\bX\trans\bX\bU - \bX\trans(\bY^t - \bZ\bA^t)\bV^t\right] + \lambda \bU(\bV^{t})\trans\bL^t\bV^t.
\end{align*}


\noindent\textbf{(b) Update of $\bV$}. 

We first reformulate the fusion penalty to simplify the update of $\bV$ under the orthogonal constraint. Note that the following equivalence holds.
\begin{proposition}\label{prop:3}
	For an arbitrary $\bB$ and membership vector $\mathbb{G}$, we have 
	\begin{align*}
		\trace{\bB\bL\bB\trans} = 2\sum_{k=1}^K \sum_{j\in G_k} \norm{\bbeta_j - \bmu_k}_2^2,
	\end{align*}
	where $\bmu_k = \abs{G_k}^{-1} \sum_{i\in G_k} \bbeta_i$, and $\bL$ is the normalized Laplacian matrix.
\end{proposition}

When updating $\bV$, other parameters should be fixed, thereby allowing us to rewrite the fusion penalty as
\begin{align*}
    \trace{\bU^{t+1}\bV\trans\bL^t\bV(\bU^{t+1})\trans}
    = 2\sum_{k = 1}^K \sum_{j\in G_k^t} \norm{\bU^{t+1}(\v_j - \m_k)}_2^2,
\end{align*}
where we utilize that $\bB=\bU^{t+1}\bV\trans$, $\v_j\in\mbR^r$ is the $j$-th row of $\bV$, and $\m_k = {\abs{G_k^t}}^{-1} \sum_{j\in G_k^t}\v_j$. Consequently, the optimization is simplified as 
\begin{equation}
    \label{V1}
    \begin{aligned}
        \min_{\bV}~
        &\frac{w_t}{2} \norm{\bY^t - \bZ\bA^t - \bX\bU^{t+1}\bV\trans}_F^2 + \lambda \sum_{k = 1}^K \sum_{j\in G_k^t} \norm{\bU^{t+1}(\v_j - \m_k)}_2^2
        \\
        &\text{s.t.} \quad \bV\trans\bV = \bI_r, \quad \m_k = \frac{1}{\abs{G_k^t}} \sum_{j\in G_k^t}\v_j,
    \end{aligned}
\end{equation}
Due to the constraint on $\m_k$ in \eqref{V1}, there is no a closed solution of \eqref{V1}. Thus we divide the problem of \eqref{V1} into two steps as follows.
\begin{algorithm}[!htp]
    \caption{Outline of Blockwise Updates}\label{alg:1}
    \begin{algorithmic}
        \State \textbf{Input}: response matrix $\bY$, predictor matrices $\bX$ and $\bZ$, cluster number $K$, sparsity level $s$ and the hyperparameter $\lambda$.
        \State \textbf{Initialization}: Let $t\leftarrow 0$, initialize $(\bA^t, \bU^t, \bV^t, \mbG^t)$ and step size $\eta$.
        \Repeat
        \State (a) Calculate the surrogate objective function in \eqref{opt:4}.
        \State (b) Update $(\bA^t, \bU^t, \bV^t, \mbG^t)$ by \eqref{update:u}, \eqref{update:v}, \eqref{update:alpha} and \eqref{update:cluster}, respectively.
        \State (c) $t\leftarrow t+1$.
        \Until{the objective function in \eqref{opt:2} converges.}
        \State \textbf{Output}: the estimation $(\bA^t, \bU^t, \bV^t, \mbG^t)$.
    \end{algorithmic}
\end{algorithm}

\noindent\underline{(b-1)}. We first set $\m_k$ as $\m_k^{t}={\abs{G_k^t}}^{-1} \sum_{j\in G_k^t}\v_j^t$, after which the optimization \eqref{V1} is simplified as 
\begin{align*}
    & \underset{\bV}{\max} ~ \trace{\bV\bS} \quad \text{s.t.} \bV\trans\bV = \bI_r,
\end{align*}
where $\bS = w_t(\bU^t)\trans\bX\trans(\bY^t - \bZ\bA^t) + 2\lambda (\bU^t)\trans\bU^t(\bM^t)\trans$. The matrix $\bM^t\in\mbR^{m\times r}$ is composed of $\{\m_k^t\}_{k=1}^K$, whose the $j$-th row is $\m_k^t$ if $j\in G_k$. This problem is a standard orthogonal procruste problem that has a closed solution
\begin{align}
    \bV^{t+1} \leftarrow \bQ_1 \bQ_2\trans, \label{update:v}
\end{align}
where $\bQ_1$ and $\bQ_2$ are the left and right singular matrices of $\bS\trans$.

\noindent\underline{(b-2)}. We then update $\m_k$ as $\m_k^{t+1} = \abs{G_k}^{-1}\sum_{j\in G_k} \v_j^{t+1}$ immediately after obtaining $\bV^{t+1}$. These two steps ensure the descent of the objective function in \eqref{V1}. 

\smallskip

\noindent \textbf{(c) Update of $\bA$, $\bM$ and $\mbG$.} 

The update of $\bA$ is simple because there is no any constraint on it. Thereby we update $\bA$ by the least squares as follows 
\begin{equation}
    \bA^{t+1} \leftarrow (\bZ\trans\bZ)^{-1}\bZ\trans(\bY^t - \bX\bB^{t+1}), 
    \label{update:alpha}
\end{equation}
with $\bB^{t+1} = \bU^{t+1} (\bV^{t+1})\trans$. The update of $\mathbb{G}$ is a $K$-means problem and we update it as follows 
\begin{equation}
    (\bM^{t+1},\mbG^{t+1}) \leftarrow \underset{\bM,\mbG}{\arg\min} ~ \sum_{k=1}^K \sum_{j\in G_k} \norm{\bU^t(\v_j^t - \m_k)}_2^2, \label{update:cluster}
\end{equation}
where we update $\bM$ once more to ensure consistency with the other parameters. 

Steps (a)–(c) constitute one complete iteration of the algorithm, which is summarized in Algorithm \ref{alg:1}. In Theorem \ref{thm:2} (Section \ref{sec:theory}), we establish that Algorithm \ref{alg:1} reduces the surrogate objective function \eqref{opt:4} monotonically, thereby guaranteeing convergence to at least a local minimizer. For further details, please refer to Theorem \ref{thm:2}.

\section{Theoretical Justifications}\label{sec:theory}

Before the main theoretical result, we first introduce some notations that will be used subsequently. For the ease of theoretical analysis, we consider the following optimization
\begin{equation}
    \begin{aligned}
         (\wh{\bA}, \wh{\bB}) = \underset{\bA,\bB}{\arg\min}\,\, \ell(\bA,\bB),\quad \text{s.t.} \,\, \norm{\bB}_{2,0} \leq s, ~ \rank{\bB} = r ~ \text{and} ~ P_{km}(\bB)\le \gamma,
    \end{aligned} \label{opt:new}
\end{equation}
where the fusion penalty $P_{km}(\bB)$ is defined as 
\begin{equation}
    \label{km}
    P_{km}(\bB)=\min_{G_1,\cdots,G_k,\bmu_1,\cdots,\bmu_K}\sum_{k=1}^K \sum_{j \in G_k} \norm{\bbeta_j - \bmu_k}_2^2.
\end{equation}
Proposition \ref{prop:3} has established that the Laplacian penalty in \eqref{opt:2} is equivalent to \eqref{km}. Furthermore, the optimization problem in \eqref{opt:2} represents the Lagrangian form of \eqref{opt:new}. Consequently, we can focus on the optimization problem in \eqref{opt:new} to investigate the theoretical properties of our estimation.

For convenience, we introduce a augmented matrix $\wt{\bB}$ to denote a joint matrix composed of $\bA$ and $\bB$, thereby the link function also is written as $b(\wt{\bB})$. Following the notation, we define the solution space of optimization \eqref{opt:new} as 
\begin{equation*}
    \wt\mbB_{rs\gamma}=\{\wt\bB=[\bA\trans,\bB\trans]\trans:\bA\in\mbR^{q\times m},\bB\in\mbR^{p\times m},\norm{\bB}_{2,0} \leq s, ~ \rank{\bB} \leq r,~ \text{and} ~ P_{km}(\bB)\le \gamma\},
\end{equation*}
which includes the true parameters that is $\wt{\bB}^*=[\bA\strans,\bB\strans]\in \wt{\mbB}_{rs\gamma}$. The parameter $\gamma$ implicitly corresponds to $\lambda$ in \eqref{opt:2} and measures the similarity level among the columns of $\bB^*$. A smaller value of $\gamma$ implies a high level of similarity and a larger $\lambda$, which encourages tighter clustering of the columns. These unknown hyperparameters $(s, r, \lambda)$ can be tuned using cross-validation in practice. Moreover, we use the notation $\wt{\bX}$ to denote the augmented covariate matrix composed of $\bZ$ and $\bX$ that is $\wt{\bX} = [\bZ,\bX]\in\mbR^{n\times (q+p)}$ then introduce the sparse operator norm of $\wt{\bX}$ as follows 
\begin{align*}
    &\norm{\wt\bX}_{s,\op} = \sup\left\{\frac{\norm{\wt\bX\v}_2}{\norm{\v}_2}:\v=[\v_1\trans,\v_2\trans]\trans\in\mbR^{q+p},\v_1\in\mbR^q, \v_2\in\mbR^p ~\text{and}~ \norm{\v_2}_0\leq s, \v\neq\0\right\},
\end{align*}
where $q$ and $p$ are dimensions of $\z_i$ and $\x_i$, respectively. Finally, we define the following notation
\begin{align*}
    \zeta_n^2 = n^{-2} \left[\min\{s\log p+m\log K,\, [s+\min(K^2,m)]\min(r,K)\log(m),\, r(s+m)\} \norm{\wt \bX}_{2s,op}^2\right],
\end{align*}
which will be used in the subsequent theorem. 

\subsection{Technical conditions}

We first introduce some technical conditions and definitions needed to clarify the theoretical result.

\begin{condition}\label{cond:1}
    The differentiable link function $b: \mbR^{(q+p)\times m} \rightarrow \mbR$ is restricted strong convex if the following holds for arbitrary $\wt\bB_1,\wt\bB_2\in \wt\mbB_{rs\gamma}$ 
    \begin{align*}
        b(\wt\bB_2) - b(\wt\bB_1) \geq \inner{\nabla b(\wt\bB_1), \wt\bB_2 - \wt\bB_1} + \frac{n\kappa}{2} \norm{\wt\bB_2 - \wt\bB_1}_F^2,
    \end{align*}
    where $\kappa$ is a positive constant.
\end{condition}

\begin{condition}\label{cond:2}
    There exists a positive constant $\rho$ satisfies that $n^{-1/2}\norm{\wt{\bX}}_{2s,\op} \leq \rho$.
\end{condition}

\begin{condition}\label{cond:4}
    Let $\bE$ be $\bY - \mbE(\bY)$ and $\e_i\in\mbR^m$ be the $i$-th row of $\bE$. Then for any given $i$, it holds that
    \begin{align*}
        \mbP( \abs{\a\trans \e_i} > \delta) \leq 2\exp\left(-\frac{\delta^2}{\tau \|\a\|_2^2}\right),
    \end{align*}
  for some constant $\tau>0$. 
\end{condition}

The restricted strong convexity in Condition \ref{cond:1} has been extensively utilized in variable selection and reduced-rank regression. For linear models, the parameter $\kappa$ in Condition \ref{cond:1} corresponds to the minimum eigenvalue of the Gram matrix restricted to $\wt{\mbB}_{rs\gamma}$, also known as the restricted eigenvalue \citep{bunea2012joint,uematsu2019sofar,chen2022fast}. In nonlinear models, $\kappa$ represents the lower bound of the eigenvalues of the Hessian matrix, a condition that is also discussed in \citet{jain2014iterative,kallus2020dynamic,wen2023simultaneous}. Analogous to Condition \ref{cond:1}, Condition \ref{cond:2} ensures smoothness within $\wt{\mbB}_{rs\gamma}$, referred to as the restricted maximum eigenvalue or restricted strong smoothness in both linear and nonlinear models \citep{uematsu2019sofar,jain2014iterative,zheng2019scalable}. Condition \ref{cond:4} assumes that the noise follows a sub-Gaussian distribution, characterized by an exponentially decaying tail probability. This encompasses a broad spectrum of distributions, including normal, Bernoulli, and multinomial distributions, thereby enabling our theoretical analysis to be applicable to generalized linear models.

\begin{theorem}\label{thm:1}
    Under Conditions \ref{cond:1} -- \ref{cond:4} and by defining $\xi_{m,K}(\gamma)$ as follows 
    \begin{align*}
        \xi_{m,K}(\gamma) = \sqrt{r\gamma(s+m)} + \sqrt{s\log p+m\log K} + \sqrt{[s+\min(K^2,m)]\min(r,K)\log m},
    \end{align*}
    we have that 
 \begin{enumerate}
     \item[(a).] If  $\frac{\kappa^2 \gamma}{\zeta_n^2}=O(1)$ as $n\to\infty$, we have 
     \begin{align*}
            \norm{\wh\bA-\bA^*}_F+\norm{\wh\bB-\bB^*}_F = O_P\left(
                \frac{\rho\xi_{m,K}(\gamma)}{\kappa\sqrt{n}} 
                \right);
     \end{align*}
     \item[(b).] If $\frac{\kappa^2 \gamma}{\zeta_n^2}\to\infty$ as $n\to\infty$, we have 
     \begin{align*}
        \norm{\wh\bA-\bA^*}_F+\norm{\wh\bB-\bB^*}_F = O_P\left(
        \frac{\rho\xi_{m,K}(1)}{\kappa\sqrt{n}} 
        \right).
     \end{align*}
 \end{enumerate}
 Consequently, the $\xi_{m, K}(\gamma)/\sqrt{n}$ concludes the statistical convergence rate of estimation.
\end{theorem}

Theorem \ref{thm:1} suggests that for the fusion penalty to work for a better convergence rate, in which the group number $K$ and the difference within each group $\gamma$ are pivotal. One hand, a small enough $\gamma$ can shrink the estimation error and the magnitude of $\gamma$ is measured by $\zeta_n^2$, where the difference within each group is helpful if $\gamma=O(\zeta_n^2)$, otherwise, the effect of $\gamma$ is invalid that is reflected by setting $\gamma=1$ in the item (b) of Theorem \ref{thm:1}. In addition, relative to the rank $r$ and the dimension $m$, a small group number $K$ also can decrease the estimation error, which corresponds to the last two terms in $\xi_{m,K}(\gamma)$, because a small $K$ will reduce the freedom degree of model. Once the group structure is indistinctive that means a too large $\gamma$ and $K$, the error will degrade into $O(\sqrt{s\log p} + \sqrt{r(s+m)\log m})$, which is same to the result of sparse reduced-rank regression excluding a logarithmic factor $\log m$ \citep{she2017selective,bunea2012joint}. Oppositely, the group structure can decrease the error mostly, where $\gamma=0$ and $K=1$. It means that there only is one group and all columns are equal mutually. Under this specific case, the estimation error is  $O_P(\sqrt{s\log(p\vee m)})$ that has degrade into the statistical accuracy of the univariate-response regression in high dimensions \citep{fan2014asymptotic,ye2010rate}. That is reasonable because the degree of freedom of coefficient is same to the univariate-response regression under the specific homogeneity, where we can vectorize the response matrix and reformulate the model as the univariate-response regression. As a summary, the statistical convergence rate presented in Theorem \ref{thm:1} concludes the theoretical results between the sparse reduced-rank regression and the univariate-response regression, which is owing to the consideration of the homogeneity. Consequently, our theoretical analysis provides a comprehensive benchmark to unify the sparsity, low rank, and homogeneity assumptions. Finally, we provide an algorithm analysis that guarantees the convergence of blockwise updates in Algorithm \ref{alg:1}, which stated as follows.
\begin{theorem}\label{thm:2}
    Following the update rules of Algorithm \ref{alg:1} and Condition \ref{cond:1}, we have 
    \begin{align*}
        \mcL(\bA^t, \bU^t, \bV^t, \bM^t, \mbG^t) \geq \mcL(\bA^{t+1}, \bU^{t+1}, \bV^{t+1}, \bM^{t+1}, \mbG^{t+1}),
    \end{align*}
    where the step size $\eta$ in the update of $\bU$ follows $\eta = O\left(n\kappa/(\norm{\bX}_{2s,\op} + \lambda)\right)$. 
\end{theorem}
The convergence of blockwise updates is ensured by the fact that each update reduces the objective loss. In step (a), the update of $\bU$ leverages iterative hard thresholding, which has been shown to decrease the loss at each iteration \citep{jain2014iterative}. In step (b), the update of 
$\bV$ reduces to an orthogonal Procrustes problem, allowing the objective function to be minimized using a closed-form solution \citep{gower2004procrustes}. For step (c), the loss reduction is achieved through the 
$K$-means and least squares procedures. As a result, under the majorization-minimization framework, the proposed algorithm converges to a critical point.

\section{Numerical Studies}\label{sec:simulation}

In the numerical simulations, we compare the performance of sparse reduced-rank regression with and without the fusion penalty, referred to as Laplacian and SRRR, respectively. Additionally, we evaluate the performance of the group Lasso estimation denoted by gLasso, which disregards low-rank and group structures and focuses solely on achieving row-sparsity. The gLasso implementation utilizes the R package \code{glmnet} \citep{simon2013blockwise}.

\subsection{General Setups}

We first outline the general setup for coefficients and covariates. In our simulations, we consider two types of covariates: one denoted by $\z_i$, whose coefficients do not exhibit low-rank or group structures; and another denoted by $\x_i$, whose coefficients possess both low-rank and group structures. Recall that the dimensions of $\z_i$ and $\x_i$ are $q$ and $p$, respectively. In the subsequent simulations, we fix $q=1$ for the low-dimensional $\z_i$. We then evaluate performance with $p=10$ and $p=100$ to represent low- and high-dimensional scenarios, respectively. Both $\z_i$ and $\x_i$ are i.i.d. drawn from a normal distribution $N(\0, \bSigma)$, where the entries of $\bSigma = (\sigma_{j,j^{\prime}})$ satisfy $\sigma_{j,j^{\prime}} = 0.5^{|j-j^{\prime}|}$.

Regarding the coefficient matrix, we primarily focus on the coefficients of $\x_i$, denoted by $\bB^*$. Let csize represent the size of each group. To construct the coefficient matrix $\bB^* \in \mbR^{p\times m}$, we first generate an $m \times r^*$ matrix $\wt{\bV}$, where $m = K \times \text{csize}$. Throughout this paper, the group size is set to $10$. For the matrix $\wt{\bV}$, its rows are divided into $K$ segments, with each segment containing csize consecutive rows of $\wt{\bV}$. The rows within the $k$-th segment, for $k=1,\dots,K$, are i.i.d. drawn from the normal distribution $N(\c_k,\epsilon^2\bI_r)$, where $\c_k = (-1)^{k+1} \lfloor (k+1)/2 \rfloor \delta \times \1$, and $\delta=3$.

Let $s^*$ denote the true sparsity level. After obtaining the matrix $\wt{\bV}$, we construct an $s^* \times s^*$ matrix $\bQ$ with entries i.i.d. drawn from $N(0, 1)$. We then extract its first $r$ left singular vectors to form a column-orthogonal matrix $\wt{\bU}^{\prime} \in \mbR^{s^* \times r}$. Let $\wt{\bU} \in \mbR^{p \times r}$ be the concatenated matrix $[\wt{\bU}\trans,\0_{(p-s^*)\times r^*}\trans]\trans$. Consequently, we obtain the matrix $\wt{\bB} = \wt{\bU}\wt{\bV}\trans$. For the coefficients of $\z_i$, we generate a matrix $\wt{\bA}^{\prime}$ with entries i.i.d. drawn from the normal distribution $N(0, 1)$. Given $\wt{\bB}$ and $\wt{\bA}^{\prime}$, we normalize these matrices as follows to derive the final coefficients.

Note that we still need to set the intercept term. For the multinomial logistic model, there are $(m+1)$ categories denoted by $\rho_j$, $j = 0, 1, \dots, m$, where $\rho_0 = 1$ represents the reference category. For $j=1,\dots,m$, the probability of category $j$ is given by $\wt{\rho}_j=\exp(\wt{\alpha}_{0j} + \z_i\trans\wt{\balpha}_j + \x_i\trans\wt{\bbeta}_j)$, where $\wt{\balpha}_j$ and $\wt{\bbeta}_j$ are the $j$-th columns of $\wt{\bA}$ and $\wt{\bB}$, respectively. The intercept $\wt{\alpha}_{0j}$ is determined as $\wt{\alpha}_{0j} = \log(\rho_0) - \bar{\z}\trans\wt{\balpha}_j - \bar{\x}\trans\wt{\bbeta}_j$, with $\bar{\x}$ being the sample mean of the covariates $\x_i$. This adjustment ensures that the intensity of each category remains close to that of the reference category. Consequently, we obtain the intercept vector $\wt{\balpha}_0 = (\wt{\alpha}_{0j})_{j=1}^m \in \mbR^m$. Define the joint matrix $\bTheta = [\wt{\balpha}_0, \wt{\bA}\trans, \wt{\bB}\trans]\trans \in \mbR^{(p+q+1)\times m}$ and let $d_{\max}$ be its maximum singular value. To control the signal strength, we normalize $\wt{\balpha}_0$, $\wt{\bA}$, and $\wt{\bB}$, resulting in the final coefficients: $\bA^* = d[\wt{\balpha}_0, \wt{\bA}\trans]\trans / d_{\max}$ and $\bB^* = d\wt{\bB} / d_{\max}$, where $d=2$ throughout this paper. Here, we combine $\wt{\balpha}_0$ and $\wt{\bA}$ in $\bA^*$ for clarity.

\subsection{Performance measure and hyperparameters tuning} 

For the linear model, we focus on the estimation error and prediction error, which are defined as follows
\begin{align*}
    \text{Err} = \norm{\wh{\bA} - \bA^*}_F^2 + \norm{\wh{\bB} - \bB^*}_F^2, ~ 
    \text{Prediction} = \norm{\bZ(\wh{\bA} - \bA^*) + \bX(\wh{\bB} - \bB^*)}_F^2.
\end{align*}
For the multinomial logistic model, we replace the prediction error with KL divergence relative to the true multinomial distribution, which is defined as follows
\begin{align*}
    \text{KL} = \frac{1}{n} \sum_{i=1}^n \sum_{j=0}^{m} \wh{P}(h_i = j) \log\frac{\wh{P}(h_i = j)}{P(h_i=j)}
\end{align*}
where $h_i$ is the category label of the $i$-th observation, $P(h_i = j)$ is the probability that the $i$-th observation belongs to category $j$, and $\wh{P}(y_i=j)$ is the estimate of $P(y_i=j)$. Specifically, $P(h_i = j)$ is given by $\exp(\wt{\z}_i\trans\balpha_j^* + \x_i\trans\bbeta_j^*) / [\sum_{j=0}^m \exp(\wt{\z}_i\trans\balpha_j^*+ \x_i\trans\bbeta_j^*)]$, where $\wt{\z}_i = (1, \z_i\trans)\trans$ is the augmented covariate vector including the intercept, and $\balpha_j^*$ and $\bbeta_j^*$ are the $j$-th columns of $\bA^*$ and $\bB^*$, respectively. Furthermore, we replace the true coefficients with their estimates to obtain $\wh{P}(h_i=j)$.

We tune the hyperparameters of each method using cross-validation (CV). For gLasso, CV is implemented via the R package \code{glmnet}. SRRR requires tuning of the rank $r$ and sparsity level $s$; therefore, we select these two hyperparameters from their crossed grid, resulting in the estimated values $\wh{s}$ and $\wh{r}$. For Laplacian SRRR, we tune the group number $K$ and penalty level $\lambda$ in their crossed grid, setting the rank and sparsity level to the previously obtained $\hat{s}$ and $\hat{r}$ from the CV of SRRR. The candidate sets for rank and sparsity level are $\{2,3,\dots,10\}$ and $\{10,11,\dots,20\}$, respectively. The candidate sets for $K$ and $\lambda$ are $\{2,3,\dots,10\}$ and $\text{seq}_{\text{e}}(-2\log 10, 0, 50)$, respectively. Here, $\text{seq}_{\text{e}}(a, b, c)$ generates an equally spaced sequence from $a$ to $b$ with length $c$, followed by taking the exponential of each element. At each point in the CV grid, we evaluate the loss function and select the hyperparameters that yield the best performance.

\subsection{Study for Linear Model}

We follow the same experimental setups as in the study of the multinomial logistic model to generate the coefficient matrices $\bA^*$ and $\bB^*$. However, the components of the intercept coefficients $\wt{\balpha}_0$ are i.i.d. drawn from the uniform distribution on $[-1, -0.5] \cup [0.5, 1]$. Given the covariates and the generated coefficients, the responses are then subsequently generated as follows
\begin{align*}
    \bY = \bZ\bA^* + \bX\bB^* + \bE,
\end{align*}
where $\bE$ is a noise matrix, its entries are i.i.d. drawn from $N(0,\sigma^2)$ and $\sigma$ is determined by the signal-to-noise rate (SNR) defined as $\text{SNR} = {\norm{\bZ\bA^* + \bX\bB^*}_F}/{\norm{\bE}_F}$. Here, $\bZ$ includes an intercept column as its first column, and we set SNR to $3$ throughout the numerical simulations. The sample size is varied with $n = 100, 150, 200$. 

\begin{table}[!htp]
    \caption{Nonsparse case of linear model: $p = s = 10$.\label{tab:linear:1}}
    \centering
    \resizebox{\textwidth}{!}{
        \subfloat[rank$=3,\, K=3$]{\begin{tabular}{cccc||cc||cc} 
\toprule 
$\epsilon$ & Method & Er$(\bB) \times 10^2$ & Prediction  & Er$(\bB) \times 10^2$ & Prediction  & Er$(\bB) \times 10^2$ & Prediction 
 \\ \hline 
 &&\multicolumn{2}{c}{$n=100$} & \multicolumn{2}{c}{$n=150$} & \multicolumn{2}{c}{$n=200$}
 \\ 
\hline 
\multirow{3}{*}{$1.8$} & gLasso &  17.12 &  9.05 &  9.65 &  8.96 &  7.8 &  9.65
 \\ 
  & SRRR &  6.58 &  4.47 &  3.86 &  4.33 &  3.1 &  4.63
 \\ 
  & Laplacian &  6.36 &  4.38 &  3.86 &  4.32 &  3.07 &  4.6
 \\ 
 \hline 
\multirow{3}{*}{$0.6$} & gLasso &  15.57 &  8.23 &  8.87 &  8.24 &  7.75 &  9.59
 \\ 
  & SRRR &  6.28 &  4.22 &  4.15 &  4.42 &  3.56 &  5.06
 \\ 
  & Laplacian &  5.32 &  3.78 &  3.62 &  4.02 &  3.11 &  4.64
 \\ 
 \hline 
\multirow{3}{*}{$0.2$} & gLasso &  15.6 &  8.24 &  8.93 &  8.3 &  7.78 &  9.62
 \\ 
  & SRRR &  5.8 &  3.98 &  3.37 &  3.89 &  2.8 &  4.46
 \\ 
  & Laplacian &  3.24 &  2.68 &  2.03 &  2.71 &  1.76 &  3.17
 \\ 
 \bottomrule 
 \end{tabular}}
        \hspace{0.2cm}
        \subfloat[rank$=6,\, K=3$]{\begin{tabular}{cc||cc||cc} 
\toprule 
 Er$(\bB) \times 10^2$ & Prediction  & Er$(\bB) \times 10^2$ & Prediction  & Er$(\bB) \times 10^2$ & Prediction 
 \\ \hline 
\multicolumn{2}{c}{$n=100$} & \multicolumn{2}{c}{$n=150$} & \multicolumn{2}{c}{$n=200$}
 \\ 
\hline 
  18.97 &  10.03 &  11.13 &  10.44 &  8.1 &  10.12
 \\ 
 12.2 &  7.66 &  7.35 &  7.82 &  5.41 &  7.53
 \\ 
 11.01 &  7.2 &  7 &  7.54 &  5.19 &  7.3
 \\ 
 \hline 
  19.17 &  10.13 &  10.7 &  10.04 &  8 &  10
 \\ 
 9.63 &  7.1 &  7.08 &  7.77 &  5.4 &  7.7
 \\ 
 7.57 &  6 &  5.28 &  6.2 &  4.22 &  6.31
 \\ 
 \hline 
  18.82 &  9.95 &  10.47 &  9.83 &  7.85 &  9.8
 \\ 
 6.49 &  4.75 &  4.19 &  4.95 &  3.13 &  5.01
 \\ 
 3.6 &  3.26 &  2.5 &  3.41 &  1.97 &  3.57
 \\ 
 \bottomrule 
 \end{tabular}}
    } 
\end{table}

\begin{table}[!htp]
    \caption{Sparse case of linear model: $p = 100, s = 10$.\label{tab:linear:2}}
    \centering
    \resizebox{\textwidth}{!}{
        \subfloat[rank$=3,\, K=3$]{\begin{tabular}{cccc||cc||cc} 
\toprule 
$\epsilon$ & Method & Er$(\bB) \times 10^2$ & Prediction  & Er$(\bB) \times 10^2$ & Prediction  & Er$(\bB) \times 10^2$ & Prediction 
 \\ \hline 
 &&\multicolumn{2}{c}{$n=100$} & \multicolumn{2}{c}{$n=150$} & \multicolumn{2}{c}{$n=200$}
 \\ 
\hline 
\multirow{3}{*}{$1.8$} & gLasso &  24.15 &  16.79 &  18.2 &  15.18 &  11.94 &  13.95
 \\ 
  & SRRR &  8.35 &  6.9 &  4.35 &  4.9 &  2.77 &  4.32
 \\ 
  & Laplacian &  8.27 &  6.78 &  4.29 &  4.84 &  2.73 &  4.27
 \\ 
 \hline 
\multirow{3}{*}{$0.6$} & gLasso &  17.63 &  13.78 &  13.18 &  12.46 &  9.47 &  11.77
 \\ 
  & SRRR &  8.65 &  7.14 &  4.9 &  5.29 &  3.27 &  4.8
 \\ 
  & Laplacian &  8.43 &  6.3 &  4.85 &  5.05 &  2.72 &  4.17
 \\ 
 \hline 
\multirow{3}{*}{$0.2$} & gLasso &  15.04 &  12.98 &  11.17 &  11.56 &  8.68 &  11.43
 \\ 
  & SRRR &  8.83 &  6.85 &  4.13 &  4.89 &  2.82 &  4.46
 \\ 
  & Laplacian &  6.21 &  4.86 &  3.67 &  3.91 &  1.75 &  2.99
 \\ 
 \bottomrule 
 \end{tabular}}
        \hspace{0.2cm}
        \subfloat[rank$=6,\, K=3$]{\begin{tabular}{cc||cc||cc} 
\toprule 
 Er$(\bB) \times 10^2$ & Prediction  & Er$(\bB) \times 10^2$ & Prediction  & Er$(\bB) \times 10^2$ & Prediction 
 \\ \hline 
\multicolumn{2}{c}{$n=100$} & \multicolumn{2}{c}{$n=150$} & \multicolumn{2}{c}{$n=200$}
 \\ 
\hline 
 23.16 &  15.37 &  15.97 &  16.24 &  10.61 &  14.09
 \\ 
 14.79 &  9.68 &  7.92 &  8.55 &  4.68 &  6.83
 \\ 
 14.33 &  9.48 &  7.36 &  8.13 &  4.45 &  6.63
 \\ 
 \hline 
 15.74 &  12.2 &  10.91 &  12.79 &  8.13 &  11.34
 \\ 
 11.65 &  9.01 &  7.42 &  9.03 &  4.84 &  7.18
 \\ 
 8.36 &  7.06 &  5.93 &  7.56 &  4.11 &  6.3
 \\ 
 \hline 
 14.3 &  11.45 &  9.7 &  11.89 &  7.21 &  10.41
 \\ 
 8.31 &  6.42 &  4.7 &  6.17 &  2.99 &  5.18
 \\ 
 4.46 &  4.1 &  2.71 &  4.01 &  1.92 &  3.61
 \\ 
 \bottomrule 
 \end{tabular}}
    }
\end{table}

The estimation accuracy of the Laplacian method is similar to that of SRRR when the within-class difference between columns in $\bB^*$ is large. Here, the within-class difference corresponds to the parameter $\epsilon$ as shown in Tables \ref{tab:linear:1} and \ref{tab:linear:2}. A smaller $\epsilon$ indicates a higher similarity among the columns within a group. As $\epsilon$ decreases, the performance of the Laplacian method improves significantly. The poor performance of gLasso can be attributed to its failure to account for both low rank and homogeneity simultaneously. In addition, when comparing Laplacian and SRRR under a small sample size, the difference in estimation accuracy is more pronounced when $K < \text{rank}$ compared to when $K = \text{rank}$. This suggests that a smaller group number can enhance estimation accuracy. Consequently, the numerical results in both nonsparse and sparse scenarios are consistent with Theorem \ref{thm:1}, where the estimation error is governed by the similarity parameter $\gamma$, the rank $r$, and the group number $K$.


\subsection{Study for Multinomial Logistic Model}

To ensure that each category has a sufficient number of observations, we define the ratio of the observation number per category as $n_{\text{ratio}} = n / m$ and set $n_{\text{ratio}}$ to 100, 150, and 200 in this simulation. Given the covariates and coefficients $(\bA^*, \bB^*)$, we generate the response vectors from the multinomial distribution.

The simulation results for the multinomial logistic model are presented in Tables \ref{tab:logit:1} and \ref{tab:logit:2}. The performance of the Laplacian SRRR method is superior to that of other methods. Specifically, the error of Laplacian SRRR decreases as in-group similarity increases, highlighting the importance of homogeneity. Additionally, the proposed method performs well in both low- and high-dimensional setups.

\begin{table}[!htp]
    \caption{Nonsparse case of the multinomial logistic model: $p = s = 10$.\label{tab:logit:1}}
    \centering
    \resizebox{\textwidth}{!}{
        \subfloat[rank$=3,\, K=3$]{\begin{tabular}{cccc||cc||cc} 
\toprule 
$\epsilon$ & Method & Er$(\bB)$ & KL $\times 10^2$ & Er$(\bB)$ & KL $\times 10^2$ & Er$(\bB)$ & KL $\times 10^2$
 \\ \hline 
 &&\multicolumn{2}{c}{$n_{ratio}=100$} & \multicolumn{2}{c}{$n_{ratio}=150$} & \multicolumn{2}{c}{$n_{ratio}=200$}
 \\ 
\hline 
\multirow{3}{*}{$1.8$} & gLasso &  4.31 &  9.03 &  4.37 &  8.89 &  4.29 &  8.66
 \\ 
  & SRRR &  3.22 &  2.65 &  2.21 &  1.85 &  1.66 &  1.47
 \\ 
  & Laplacian &  2.7 &  2.08 &  2.12 &  1.72 &  1.57 &  1.3
 \\ 
 \hline 
\multirow{3}{*}{$0.6$} & gLasso &  3.97 &  8.8 &  3.97 &  8.7 &  3.98 &  8.47
 \\ 
  & SRRR &  2.83 &  2.58 &  1.84 &  1.66 &  1.45 &  1.26
 \\ 
  & Laplacian &  2.25 &  1.81 &  1.59 &  1.18 &  1.31 &  0.91
 \\ 
 \hline 
\multirow{3}{*}{$0.2$} & gLasso &  3.92 &  8.63 &  3.93 &  8.53 &  3.93 &  8.31
 \\ 
  & SRRR &  2.8 &  2.53 &  1.84 &  1.63 &  1.38 &  1.24
 \\ 
  & Laplacian &  2.21 &  1.75 &  1.56 &  1.1 &  1.17 &  0.79
 \\ 
 \bottomrule 
 \end{tabular}}
        \hspace{0.2cm}
        \subfloat[rank$=6,\, K=3$]{\begin{tabular}{cc||cc||cc} 
\toprule 
Er$(\bB)$ & KL $\times 10^2$ & Er$(\bB)$ & KL $\times 10^2$ & Er$(\bB)$ & KL $\times 10^2$
 \\ \hline 
\multicolumn{2}{c}{$n_{ratio}=100$} & \multicolumn{2}{c}{$n_{ratio}=150$} & \multicolumn{2}{c}{$n_{ratio}=200$}
 \\ 
\hline 
 4.33 &  8.87 &  4.46 &  8.95 &  4.47 &  9.01
 \\ 
  3.16 &  2.99 &  2.4 &  2.19 &  1.91 &  1.75
 \\ 
 2.65 &  2.44 &  2.18 &  1.88 &  1.81 &  1.55
 \\ 
 \hline 
 3.94 &  8.54 &  3.93 &  8.33 &  3.94 &  7.98
 \\ 
  2.76 &  2.51 &  1.76 &  1.69 &  1.4 &  1.27
 \\ 
 2.24 &  1.8 &  1.51 &  1.18 &  1.21 &  0.86
 \\ 
 \hline 
 3.9 &  8.38 &  3.89 &  8.19 &  3.89 &  7.83
 \\ 
  2.81 &  2.51 &  1.79 &  1.65 &  1.3 &  1.25
 \\ 
 2.25 &  1.72 &  1.53 &  1.11 &  1.09 &  0.81
 \\ 
 \bottomrule 
 \end{tabular}}
    }
\end{table}

\begin{table}[!htp]
    \caption{Sparse case of the multinomial logistic model: $p = 100, s = 10$.\label{tab:logit:2}}
    \centering
    \resizebox{\textwidth}{!}{
        \subfloat[rank$=3,\, K=3$]{\begin{tabular}{cccc||cc||cc} 
\toprule 
$\epsilon$ & Method & Er$(\bB)$ & KL $\times 10^2$ & Er$(\bB)$ & KL $\times 10^2$ & Er$(\bB)$ & KL $\times 10^2$
 \\ \hline 
 &&\multicolumn{2}{c}{$n_{ratio}=100$} & \multicolumn{2}{c}{$n_{ratio}=150$} & \multicolumn{2}{c}{$n_{ratio}=200$}
 \\ 
\hline 
\multirow{3}{*}{$1.8$} & gLasso &  4.44 &  9.31 &  4.38 &  8.29 &  4.4 &  9.06
 \\ 
  & SRRR &  4.75 &  4.12 &  2.99 &  2.59 &  2.24 &  1.94
 \\ 
  & Laplacian &  2.9 &  2.69 &  2.23 &  1.95 &  1.76 &  1.51
 \\ 
 \hline 
\multirow{3}{*}{$0.6$} & gLasso &  4.04 &  8.28 &  4.04 &  7.98 &  4.04 &  8.22
 \\ 
  & SRRR &  4.16 &  3.62 &  2.74 &  2.33 &  1.92 &  1.73
 \\ 
  & Laplacian &  2.42 &  2.12 &  1.78 &  1.4 &  1.38 &  1.06
 \\ 
 \hline 
\multirow{3}{*}{$0.2$} & gLasso &  3.99 &  8.12 &  3.99 &  7.83 &  3.99 &  8.06
 \\ 
  & SRRR &  4.05 &  3.57 &  2.61 &  2.26 &  1.85 &  1.66
 \\ 
  & Laplacian &  2.38 &  2.01 &  1.72 &  1.32 &  1.27 &  0.92
 \\ 
 \bottomrule 
 \end{tabular}}
        \hspace{0.2cm}
        \subfloat[rank$=6,\, K=3$]{\begin{tabular}{cc||cc||cc} 
\toprule 
Er$(\bB)$ & KL $\times 10^2$ & Er$(\bB)$ & KL $\times 10^2$ & Er$(\bB)$ & KL $\times 10^2$
 \\ \hline 
\multicolumn{2}{c}{$n_{ratio}=100$} & \multicolumn{2}{c}{$n_{ratio}=150$} & \multicolumn{2}{c}{$n_{ratio}=200$}
 \\ 
\hline 
 4.56 &  8.55 &  4.49 &  8.34 &  4.55 &  8.83
 \\ 
 4.69 &  4.13 &  3.08 &  2.82 &  2.34 &  2.25
 \\ 
 3.13 &  3.06 &  2.33 &  2.25 &  1.96 &  1.88
 \\ 
 \hline 
 4.03 &  7.94 &  4.01 &  7.96 &  4.01 &  8.09
 \\ 
 4.26 &  3.6 &  2.63 &  2.36 &  1.96 &  1.78
 \\ 
 2.55 &  2.2 &  1.87 &  1.55 &  1.47 &  1.2
 \\ 
 \hline 
 3.98 &  7.79 &  3.96 &  7.8 &  3.96 &  7.94
 \\ 
 3.9 &  3.5 &  2.53 &  2.29 &  1.89 &  1.72
 \\ 
 2.37 &  2.08 &  1.73 &  1.45 &  1.37 &  1.07
 \\ 
 \bottomrule 
 \end{tabular}}
    }
\end{table}

Similar to the performance under the linear model, the estimation accuracy of the Laplacian method is comparable to that of SRRR with large $\epsilon$. However, the accuracy of the Laplacian method surpasses that of SRRR when $\epsilon$ is small. Additionally, in high-dimensional cases with a small number of observations, the performance of SRRR does not outperform gLasso, which can be attributed to the limited observations within each category. Consequently, the variance associated with high-dimensional responses is larger compared to that under the linear model. Both the $\ell_2$-norm penalty and the group Lasso penalty help reduce variance and balance variance and bias, leading to a lower mean squared error. Therefore, gLasso performs better than SRRR when the number of observations is small. As the number of observations increases, the accuracy of SRRR improves and surpasses that of gLasso due to its ability to exploit low-rank structure.

\subsection{Tree Species Analysis on Barro Colorado Island}\label{sec:bci}

\begin{table}[!htp]
	\centering
	\caption{Description of covariates}
	\resizebox{\textwidth}{!}{
		\begin{tabular}{c|p{8cm}||c|p{8cm}}
			\toprule
			Covariate & Description           & Covariate         & Description         \\
			\hline
			Cu        & Copper content       & mrvbf    & multiresolution index of valley bottom flatness   \\
			\hline
			convi     & convergence index with direction to the center cell   & Nmin       & mineralization needs for Nitrogen after a 30-day incubation period \\
			\hline 
			dem       & 5m resolution elevation           & P        & Phosphorus content       \\
			\hline 
			devmean   & deviation from mean value of twi    & pH                     & pH content    \\
			\hline 
			difmean   & difference from the mean value of twi & solar            & incoming mean annual solar radiation      \\
			\hline 
			grad      & slope of gradient         & twi       & topographic wetness index     \\
			\hline 
			K         & Potassium content  soil    &            &              \\
			\bottomrule
		\end{tabular}
	}
\end{table}

\begin{table}[!htp]
	\centering
	\caption{The summary of clustered species. \label{tab:real-data}}
	\resizebox{0.9\textwidth}{!}{
			\begin{tabular}{c|l}
				\toprule
				Group ID & Species                                                        \\
				\hline
				1          & ALCHCO, ZANTBE                                                 \\
				6          & CORDAL, ERY2MA, HAMPAP, LICAHY, OCOTCE, OCOTOB, OCOTPU, ZANTP1 \\
				10         & ANDIIN, ARDIFE, CELTSC, HURACR, NECTGL, ORMOCR, TROPRA         \\
				11         & APEIME, CASEAR, INGASA, JAC1CO                                 \\
				13         & CASESY, GUAPST, LACMPA, ORMOMA, PLA1PI, PLA2EL                 \\
				14         & ANNOSP, PENTMA, PIPERE, SYMPGL, VIROSU                         \\
				15         & AST1ST, ERY2PA, HEISAC, LUEHSE                                 \\
				16         & ALLOPS, CHR2CA, COCCCO, SENNDA                                 \\
				\bottomrule
			\end{tabular}%
	}
\end{table}

\begin{figure}[!htp]
	\centering
	\subfloat[Group 6]{\includegraphics[width=0.5\textwidth]{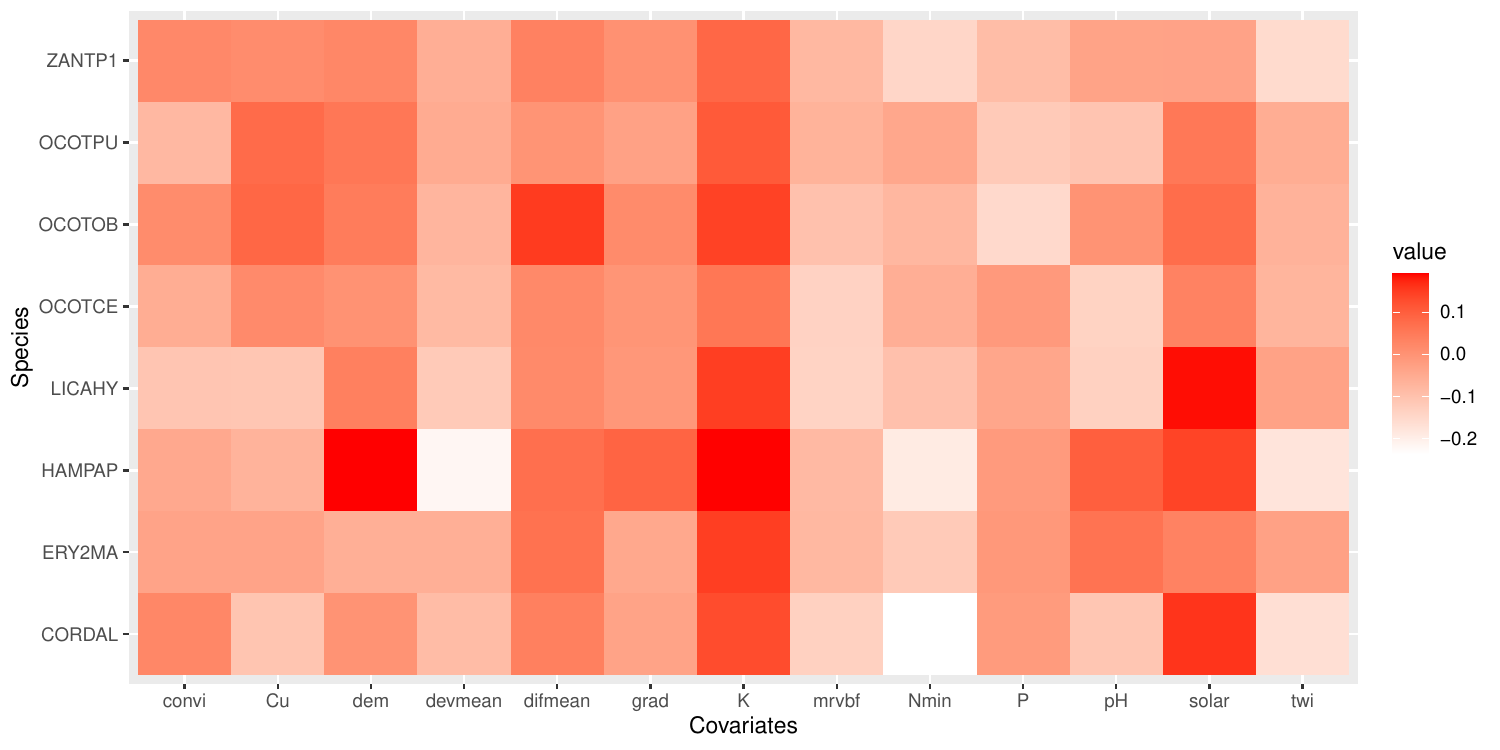}}
	\subfloat[Group 10]{\includegraphics[width=0.5\textwidth]{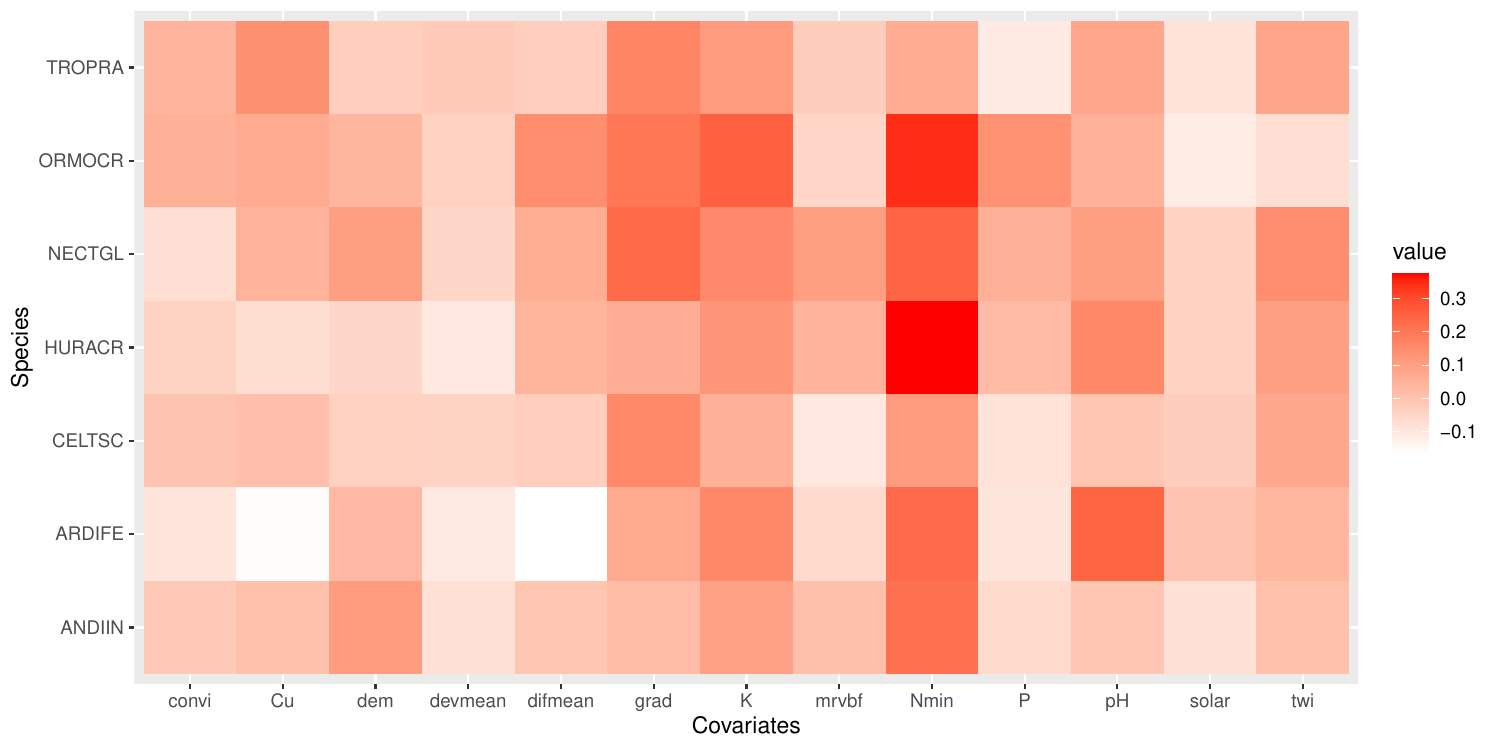}} \\
	\subfloat[Group 13]{\includegraphics[width=0.5\textwidth]{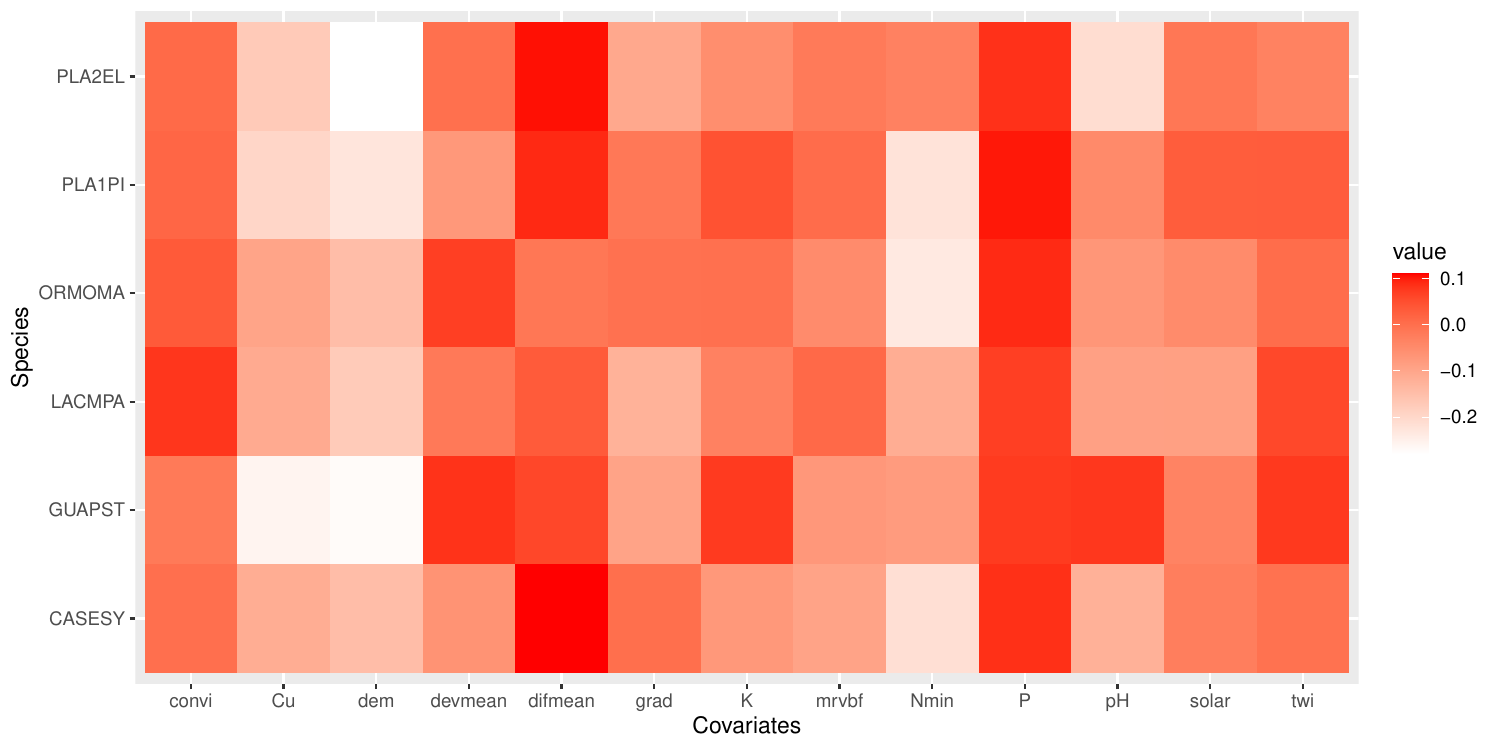}}
	\subfloat[Group 14]{\includegraphics[width=0.5\textwidth]{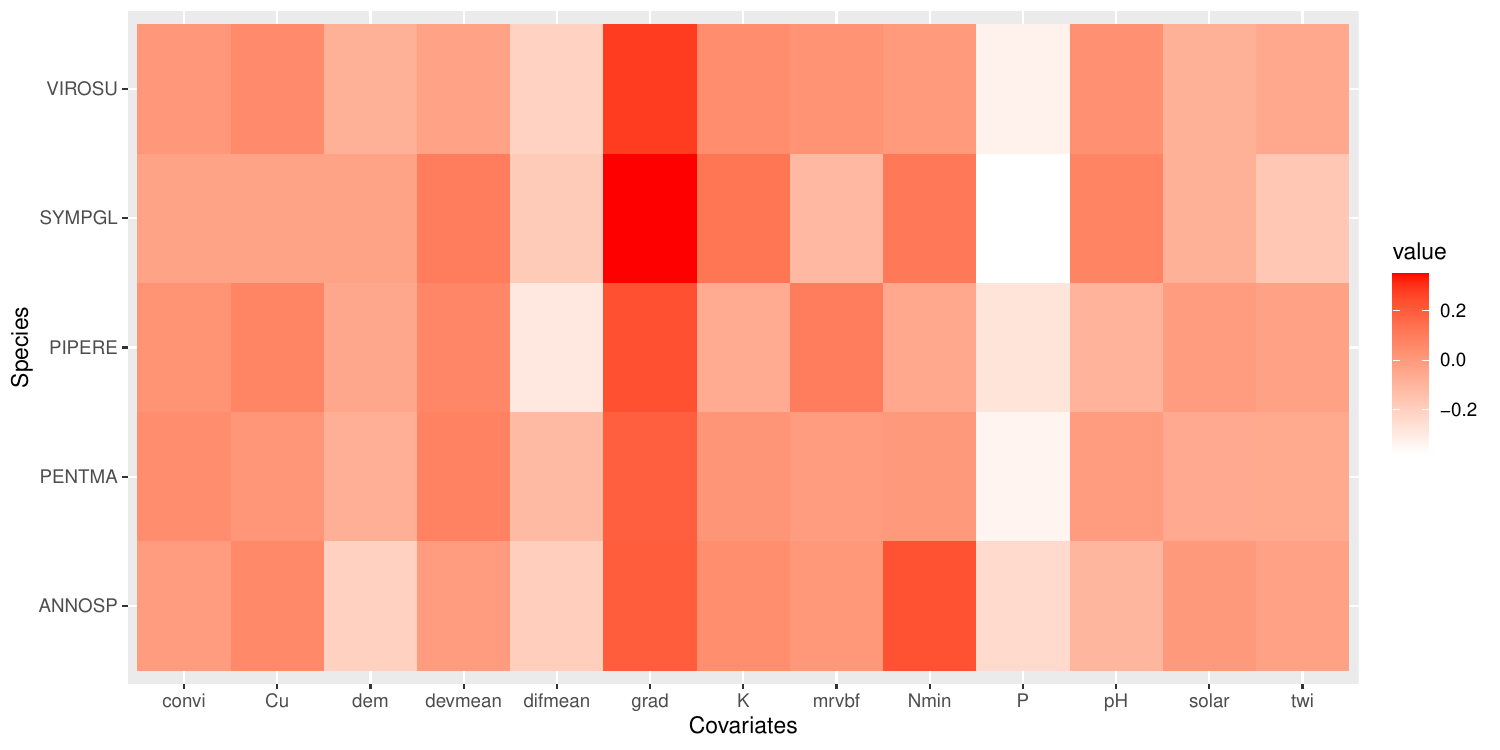}}
	\caption{Heatmaps of coefficients.\label{fig:heatmap}}
\end{figure}

The Barro Colorado Island (BCI) dataset provides information on 3000 species, comprising 350,000 trees surveyed in a rectangular area measuring 1000m by 500m. We focus on species with observation numbers between 100 and 300, extracting a subset of 51 species from the BCI data, which includes 8,777 trees denoted by points. Following the multinomial logistic model proposed by \citet{hessellund2022semiparametric}, we fit these species using the Laplacian SRRR method. The first column of the response matrix corresponding to species ADE1TR is used as the reference category for the multinomial logistic regression. 
Moreover, we assign all covariates corresponding to the $\x$ in the aforementioned model and set the intercept corresponding to $\z$. The resulting covariate matrix has a rank of 13, encompassing all 13 covariates, indicating that the coefficient matrix is neither sparse nor low-rank. However, Laplacian SRRR clusters 50 species (excluding the baseline species) into 18 distinct groups. 
Excluding groups that contain only a single species, the clustering results for the remaining groups are summarized in Table \ref{tab:real-data}.
This group structure is evident from the heatmap of coefficients, which clearly shows the clustered patterns among species. For brevity, heatmaps for the first four largest groups are presented in Figure \ref{fig:heatmap}.




For group 6, the potassium content positively influences these species. Among these species, OCOTCE, OCOTOB, and OCOTPU belong to the Ocotea genus. Besides the Ocotea genus, other species in this group also belong to the flowering plants within the same family as Ocotea. Potassium promotes flowering and fruit growth in these plants, which aligns with the positive effect shown in the heatmap. In group 10, species such as ORMOCR and ANDIIN belong to the Fabaceae:Papilionoideae family. The heatmap for group 10 indicates significant nitrogen mineralization, suggesting these species have strong nitrogen-fixing abilities. Additionally, the negative effect of difference from the mean value of topographic wetness index (difmean) in group 10 implies that these species are likely near streams, as a smaller difmean value indicates a higher probability of stream presence \citep{hengl2009practical}. In contrast, species in group 13 exhibit a positive correlation with difmean, indicating their distribution is farther from streams, consistent with the habitat preferences of PLA2EL, a species belonging to the Platypodium genus typically found in seasonally dry tropical forests \citep{platypodium}. Moreover, species in group 13 show a strong dependence on phosphorus content; ORMOMA, belonging to the Ormosia genus, has bean-type fruits that require more phosphorus for growth \citep{lynch1991vegetative}. In summary, our method can identify commonalities among different tree species, aiding in the analysis of species connections.

\section{Discussion}

In this paper, we investigate the multi-response generalized linear model under the assumptions of column homogeneity, low rank, and sparsity. While these assumptions are commonly employed in the literature, there is limited research that addresses them within a unified framework. We employ regularization techniques and matrix factorization to incorporate the homogeneity and low-rank constraints. Additionally, sparsity is enforced by limiting the number of nonzero rows in the coefficient matrix. Based on these setups, we propose a regularized maximum likelihood estimation method and an iterative block gradient descent algorithm for estimating the coefficient matrix. We then establish the statistical consistency of our estimates, elucidating how the homogeneity and low-rank structures impact estimation accuracy. This work bridges the gap between low rank and homogeneity assumptions and extends the scope of multi-response regression. Given the broad applicability of generalized linear models, our method can be applied to various fields such as operations and marketing research, including product pricing, assortment optimization, and conjoint analysis \citep{keskin2014dynamic,kallus2020dynamic,chen2017modeling}. Future work will explore these applications further. Moreover, our theoretical analysis reveals that estimation accuracy depends on the similarity level within each group. Therefore, developing a quantitative measure for the differences among columns would be both interesting and important for practical applications. Consequently, statistical inference regarding the estimation of the matrix and the differences among its columns will also be a focus of future research.

  \bibliographystyle{plainnat}
  \bibliography{ref}

\end{document}